\documentclass[conference]{IEEEtran}
\IEEEoverridecommandlockouts
\usepackage{amsmath,amssymb,amsfonts}
\usepackage{mathtools}
\usepackage{algorithmic, algorithm}
\usepackage{graphicx}
\usepackage{textcomp}
\usepackage{xcolor}
\usepackage[table]{xcolor} 
\definecolor{paramCol}{RGB}{220, 230, 241} 
\definecolor{valueCol}{RGB}{235, 245, 230} 
\def\BibTeX{{\rm B\kern-.05em{\sc i\kern-.025em b}\kern-.08em
    T\kern-.1667em\lower.7ex\hbox{E}\kern-.125emX}}

\usepackage[T1]{fontenc}
\usepackage[utf8]{inputenc}
\usepackage[caption=false]{subfig}
\usepackage{pgfplots}
\usepackage{pgfplotstable}
\usepgfplotslibrary{fillbetween}
\usepackage{grffile}
\pgfplotsset{compat=1.18}
\usetikzlibrary{plotmarks, arrows.meta}
\usepgfplotslibrary{patchplots}
\usepackage[acronym,shortcuts]{glossaries}\newacronym{awgn}{AWGN}{additive white Gaussian noise}

\newacronym{caf}{CAF}{cross ambiguity function}
\newacronym{cndr}{CNDR}{carrier-to-noise density ratio}

\newacronym{gpsdo}{GPSDO}{GPS disciplined oscillator}
\newacronym{gnss}{GNSS}{global navigation satellite system}

\newacronym{if}{IF}{intermediate frequency}

\newacronym{kf}{KF}{Kalman filter}

\newacronym{leo}{LEO}{low Earth orbit}
\newacronym{lnb}{LNB}{low-noise block}
\newacronym{lnbf}{LNBF}{low-noise block feedhorn}
\newacronym{ls}{LS}{least squares}

\newacronym{mcrb}{MCRB}{modified Cramér-Rao bounds}
\newacronym{meo}{MEO}{medium Earth orbit}

\newacronym{ocxo}{OCXO}{oven-controlled crystal oscillator}
\newacronym{ofdm}{OFDM}{orthogonal frequency division multiplexing}

\newacronym{pnt}{PNT}{positioning, navigation and timing}
\newacronym{psk}{PSK}{phase-shift keying}

\newacronym{sdr}{SDR}{software-defined radio}
\newacronym{sgp4}{SGP4}{simplified general perturbation model 4}
\newacronym{snr}{SNR}{signal-to-noise ratio}
\newacronym{sop}{SOP}{signal of opportunity}
\newacronym{sss}{SSS}{secondary synchronization sequence}
\newacronym{sv}{SV}{space vehicle}

\newacronym{tcxo}{TCXO}{temperature-controlled crystal oscillator}
\newacronym{toa}{ToA}{time of arrival}
\newacronym{tle}{TLE}{two-line element}

\newacronym{pss}{PSS}{primary synchronization sequence}
\newacronym{pvt}{PVT}{position, velocity, and time}
\usepackage{soul}
\sethlcolor{yellow}
\usepackage{bm}
\usepackage{comment}
\usepackage{siunitx}
\usepackage{multirow}
\usepackage{cite}
\usepackage{nicefrac}
\usepackage{xfrac}
\usepackage{balance}

\newlength\fheight 
\newlength\fwidth 
\begin{document}

\title{Lightweight Pilot Estimation on LEO Satellite
Signals for Enhanced \acrshort{sop} Navigation}

\author{ \IEEEauthorblockA{Francesco~Zanirato$^{1\star}$, Alessio Curzio$^{2}$, Francesco Ardizzon$^{1}$, Elisa Sbalchiero$^{2}$, \\ Luca Canzian$^2$, Stefano Tomasin$^{1}$, Nicola Laurenti$^1$, Jaron Samson$^3$\\ 
}\vspace{2mm}
 \IEEEauthorblockA{$^1$ Department of Information Engineering, University of Padova, Italy\\
 $^2$ Qascom S.r.l., Cassola, Vicenza, Italy\\ 
 $^3$ ESTEC, European Space Agency, Noordwijk, the Netherlands\\\vspace{2mm}
 \small $^\star$ Corresponding author, email: francesco.zanirato@phd.unipd.it
 }\thanks{This work was partially funded by the European Space Agency under contract n. 4000143575/24/NL/WC/kg: “Navigation Using Machine lEaRning applied to Signals of Opportunity (NUMEROSO)”. 
}}

\maketitle

\begin{abstract}
The computation of \ac{pnt} via \ac{sop}, where signals originally transmitted for communication, such as 5G, Wi-Fi, or DVB-S, are exploited due to their ubiquity and spectral characteristics, is an emerging research field. 
However, relying on these signals presents challenges, including limited knowledge of the signal modulation and the need to identify recurring sequences for correlation.
We offer a guide to implement a receiver capable of capturing broadband downlink Ku-band signals from \ac{leo} satellites (e.g., Starlink and OneWeb) and estimating the recurring symbols for \ac{sop} measurements. The methodology integrates recent approaches in the literature, highlighting the most effective aspects while guiding the replication of experiments even under limitations on the front-end gain and bandwidth.
Using the proposed model, we can identify recurring symbols transmitted by Starlink satellites, which are then used to collect Doppler shift measurements over a \SI{600}{\second} interval. A \ac{pvt} solution is also computed via \ac{ls}, which achieves a positioning error of approximately \SI{268}{\meter} after a post-fit refinement.

\end{abstract}


\glsresetall

\section{Introduction}
Employing \ac{leo} satellites in navigation is a profitable strategy to improve current \ac{pnt} techniques. In fact, compared to traditional \ac{gnss}, the proximity of \ac{leo} satellites to Earth results in a stronger received power, reduced atmospheric effects, and higher geometry diversity during the same observation time, thanks to their faster dynamics. This potential can be exploited to achieve a more accurate and secure \ac{pnt} service, e.g.,~\cite{Ardizzon2025ML}.
\Ac{leo} \ac{pnt} may be provided by dedicated services or even by \ac{leo} communication systems offering \ac{pvt} as a secondary service. Nonetheless, we consider the use of signals originally designed for communications from \ac{leo} satellites as \ac{sop} for \ac{pnt}. 
 
 
In general, navigation through \acp{sop} presents several challenges compared to GNSS, e.g., lack of knowledge about signal encoding or modulation format.
In~\cite{Humphreys2023}, the authors disclosed the Starlink signal structure, characterizing the waveform and its \ac{ofdm} modulation format, which consists of 1024 subcarriers, including four gutter tones at the center of the channel, and a cyclic prefix of 32 chips. The symbols marking the beginning of each frame, \ac{pss} and \ac{sss}, were published in~\cite{komodromos2023}. 

Despite the large bandwidth, \ac{leo} \acp{sop} generally lack the synchronization needed for pseudorange estimation, as their frame timing is not precisely synchronized with a common reference~\cite{qin2025}. Thus, Doppler-based observables are the primary sources for navigation. 
On the other hand, \ac{pss} and \ac{sss} are designed to compensate for carrier frequency offset up to a small fraction of the subcarrier bandwidth, and their duration $T_{\mathrm{PSS, SSS}} = 8.8\,\mathrm{\mu s}$ every $T_{\mathrm{fr}} = 1/750\,\mathrm{s}$ is too short to achieve the measurement accuracy required for \ac{pnt}.  
A first solution could be to combine the estimates across different frames, coherently or not-coherently. However, the wide variation of the Doppler frequency shift between frames and the low duty cycle of \ac{pss} and \ac{sss} make both approaches difficult in practice.
 

Another approach to improve the accuracy is to investigate the \ac{sop} frame structure to identify the presence of undisclosed pilot symbols, that can increase the integration period. Along this line of research, \cite{kozhaya2024TRICK} shows that the Starlink signal includes a pilot symbol pattern besides \ac{pss} and \ac{sss}, named \emph{full beacon}, scattered over several \ac{ofdm} symbols and subcarriers and repeated across different frames and \acp{sv}. By using the full beacon, the correlation window is extended from $8.8\,\mu$s to $1.33\,$ms. The Cramér-Rao bound shows that this could lead to a standard deviation accuracy for Doppler frequency estimation approximately between $10$ and $100\,$Hz for a $C/{N_0}$ between $40$ and $60\,$dB-Hz~\cite{zanirato2024}. 
 
In \cite{kozhaya2022ORBCOMM}, a \ac{kf} is used to track the phase of the \ac{sop} from Orbcomm for beacon estimation. The approach was later generalized in~\cite{Kozhaya2023Multi} and was also applied to Starlink, Iridium NEXT, and OneWeb \ac{sop}. The hardware setup included a high-gain dish antenna and a receiver front-end capable of capturing the $250\,$MHz wide OneWeb band (the widest among the considered systems). 
A setup that does not include the dish antenna is considered in~\cite{mooseli2025}. Still, the authors do not aim at estimating the Starlink beacon rather at updating it starting from a previous estimate. 
It is worth mentioning that, at the time of writing, no publicly available version of these beacons exists to be used as a starting point. Moreover, the use of high-gain antennas significantly narrows the antenna field of view, which is particularly critical for \ac{leo} satellites due to their rapidly changing geometry, and is unsuitable for many applications due to the increase on the overall system bulk.

In turn, the first evidence of the feasibility of exploiting LEO beacons from scratch without a dish antenna was presented in \cite{Kozhaya2025}. However, the main focus of that work is to formally demonstrate that the proposed estimation technique converges to a single beacon independently of the number of transmitting sources and of their modulation schemes.
 
In this paper, we perform a full beacon estimation on Starlink \acp{sop} using a lightweight setup, including a \ac{lnb} with a feedhorn antenna and a front end capturing less than half the signal bandwidth and, consequently, less than half the recurrent symbols.
Thus, we show that full beacon estimation is feasible even with a low-gain antenna and at low \ac{snr}. The designed method is also for a mobile scenario, where the user may perform a new beacon estimate or update a previous, possibly outdated, one.
We remark that the approach can be generalized to other \ac{sop}, despite our focus on Starlink, motivated by the wide available literature.
Finally, we evaluate the \ac{pnt} performance achieved using the Doppler shifts acquired using the estimated beacon.
In summary, the main contributions of this paper are as follows. 
 \begin{itemize} 
    \item We present a lightweight hardware setup, compared to existing solutions~\cite{Kozhaya2023Multi, kassas_journal}, that is capable of performing both beacon estimation and \ac{pnt} for Starlink as \ac{sop}.
    \item We describe in detail the beacon estimation algorithm, aiding in the full replicability of the process and offering additional insights into the Starlink signal frame.
    \item We discuss the acquisition results of Doppler-shift measurements obtained with the estimated Starlink beacon.
    \item We describe the \ac{ls}-based navigation engine from Doppler-shift measurements for \ac{pnt}, and investigate the tradeoff between observation period and solution accuracy. 
 \end{itemize}
 The paper is structured as follows. Section~\ref{sec:setup} illustrates the signal capture setup. Section~\ref{sec:signal_model} presents the signal model. Section~\ref{sec:algorithm} provides the beacon estimation algorithm and offers further information on parameter choice and the overall procedure, with Section~\ref{sec:stralink_beacon} describing the estimation results. Section~\ref{sec:navigation} details the acquisition and navigation frameworks. Finally, Section~\ref{sec:conclusions} concludes the paper. 
 
\section{Signal Capture Setup} \label{sec:setup}

This section describes the RF front-end and recording chain used for the Ku-band captures. The setup was designed to (i) downconvert the Ku-band downlink to an \ac{if}, digitize a \SI{100}{\mega\hertz} complex bandwidth with a coherent \ac{sdr} front-end, and record raw IQ samples for offline processing.

\subsection{Receiver Chain Overview}
\begin{figure}
    \centering
    \includegraphics[width=\linewidth]{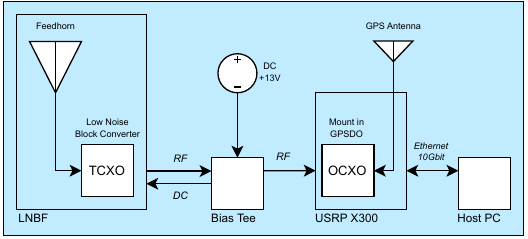}
    \caption{Receiver Chain. }
    \label{fig:receiver_chain}

    \vspace{-0.3cm}
\end{figure}

\begin{figure}
\centering
\includegraphics[width=0.8\columnwidth]{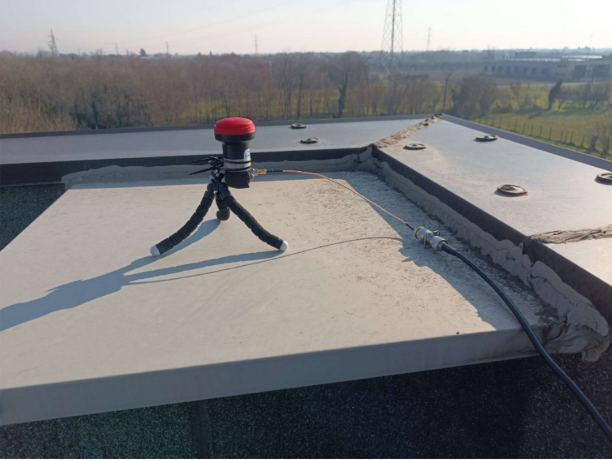}
\caption{LNB used to capture the Starlink signal placed on the rooftop of the building at the Qascom's premise.}
\label{fig:foto_setup}
\vspace{-0.4cm}
\end{figure}

An upward-pointing Ku-band receiving front-end was assembled using a commercial \ac{lnbf} (Fig.~\ref{fig:foto_setup}), powered through a bias-tee. The \ac{lnbf} downconverts the received Ku-band signal to an L-band \ac{if}. The \ac{if} output is fed to an Ettus USRP X300, which digitizes and streams complex IQ samples to a host PC for storage and offline processing. Frequency stability on the \ac{sdr} side is ensured by the \ac{gpsdo} mounted on the USRP X300, providing the \SI{10}{\mega\hertz} reference from the GPS signal to the \ac{sdr}. To simplify the RF chain, the \ac{lnbf} relies instead on its internal \ac{tcxo}. The \ac{lnbf} is linearly polarized; therefore, while the downlink is circularly polarized, a polarization mismatch loss up to \SI{3}{\decibel} is expected to be observed. Fig.~\ref{fig:receiver_chain} shows a diagram of the receiver chain.

\subsection{Recording Configuration}

The setup was configured to capture a signal centered at a carrier frequency $f_{\rm c} \approx \SI{11.325}{\giga\hertz}$, i.e., the Starlink channel 3~\cite{Humphreys2023}, with a complex bandwidth of \SI{100}{\mega\hertz}. The downconversion is performed by the \ac{lnbf} using a local oscillator at $f_{\rm LO}=\SI{9.75}{\giga\hertz}$, so that the signal given as input to the USRP is centered at an \ac{if} of $f_{\rm IF} = f_{\rm c} - f_{\rm LO} \approx \SI{1.575}{\giga\hertz}$. Finally, the sampling rate was set to $F_\mathrm{s} = \SI{100}{MS/s}$ with int16 IQ quantization. Data were recorded for \SI{600}{\second}, producing a binary file of $\approx\SI{240}{\giga\byte}$.
Table~\ref{tab:setup} summarizes the hardware setup.

\begin{table}
    \centering   
    \caption{Hardware Components used for the Setup.}
    \label{tab:setup}
    \resizebox{\columnwidth}{!}{%
    \begin{tabular}{|l|ll|}
    \hline
    \textbf{Component}       & \multicolumn{2}{p{0.7\linewidth}|}{\textbf{Details}}                                                     \\ \hline
    SDR front-end            & \multicolumn{2}{p{0.7\linewidth}|}{Ettus USRP X300 (daughterboard: UBX-160MHz); interface: 10 GbE/PCIe;} \\ \hline
    Reference \& timing      & \multicolumn{2}{p{0.7\linewidth}|}{On-board GPSDO providing $10\,$MHz ref. to SDR}                     \\ \hline
    LNBF                     & \multicolumn{2}{p{0.7\linewidth}|}{Othernet Bullseye TCXO LNBF; $f_{\rm LO}=\SI{9.75}{\giga\hertz}$, Linear Polarization}       \\ \hline
    Bias-tee                 & \multicolumn{2}{p{0.7\linewidth}|}{Model: ZFBT-282-1.5A+}                                                \\ \hline
    Cabling \& passives      & \multicolumn{2}{p{0.7\linewidth}|}{$\approx\SI{15}{\meter}$, Ultraflex 7, \SI{50}{\ohm}}                                         \\ \hline
    \multirow{5}{*}{Host PC} & \multicolumn{1}{l|}{CPU}             & Intel Core i9-13900K (24 cores, 32 threads)        \\ \cline{2-3} 
                             & \multicolumn{1}{l|}{Memory}          & 32 GB DDR5-4800 (2×16 GB)                          \\ \cline{2-3} 
                             & \multicolumn{1}{l|}{Storage}         & 2×1 TB NVMe SSD (Seagate FireCuda 530)             \\ \cline{2-3} 
                             & \multicolumn{1}{l|}{NIC}             & Intel X710-DA2 dual-port 10 GbE (2×SFP+)           \\ \cline{2-3} 
                             & \multicolumn{1}{l|}{OS}              & Windows 11 Pro (64-bit)                            \\ \hline
    \end{tabular}%
    }
\end{table}

\subsection{Relation to the State-of-the-art Setups}
The adopted chain is consistent with the widely used architecture employed in Starlink experimental studies, but differs in three key hardware aspects: captured bandwidth, reference distribution, and antenna type. Table~\ref{tab:setComp} summarizes the key hardware differences versus the literature.
\begin{table}
    \centering
    \renewcommand*{\arraystretch}{1.1}
    \caption{Key Hardware Differences Versus Literature.}
    \label{tab:setComp}
    \begin{tabular}{|p{0.25\columnwidth}|p{0.3\columnwidth}|p{0.3\columnwidth}|} \hline
    \textbf{Component} &	\textbf{Literature}	& \textbf{This work} \\ \hline 
    Capture Bandwidth	& $\geq \SI{240}{\mega\hertz}$ &	\SI{100}{\mega\hertz} \\ \hline 
    Ref. distributed to LNBF &	Yes (LNBF disciplined)	&No (LNBF TCXO only) \\ \hline 
    Antenna  &	High-gain parabolic dish ($\approx \SI{35}{dBi}$) \cite{kassas_journal} &	No dish;  upward-pointing LNBF  \\ \hline 
    \end{tabular}
    \vspace{-0.3cm}
\end{table}

In \cite{Kozhaya2025} and \cite{kassas_journal} the full \SI{240}{\mega\hertz} OFDM beacon is estimated from a \SI{1}{\second} signal recording, collected using an LNBF connected to a \SI{60}{\centi\meter} parabolic dish, and an NI USRP X410 with \SI{500}{\mega\hertz} receiver bandwidth. The setup included a common external high-stability reference within the capture chain. 
In comparison, the capture of this work adopts a reduced \SI{100}{\mega\hertz} recorded bandwidth and a TCXO-based LNBF. 
We will show that this setup has reduced data rate and storage requirements while retaining bandwidth sufficient for partial beacon estimation despite the larger Starlink OFDM beacon channel (\SI{250}{\mega\hertz}).


\section{Signal Model} \label{sec:signal_model}

Here, we describe the signal processing block, which was inspired by~\cite{Kozhaya2023Multi}.
The received downlink signal, in baseband, is constituted by a sequence of frames, each transmitted by a single \ac{sv} and characterized by its own code phase and Doppler frequency shift, which is affected by its motion. Furthermore, we can distinguish between data frames, collecting variable and unpredictable information symbols, and pilots, which are deterministic and assumed to be constant across different frames and satellites. The deterministic part forms the full beacon~\cite{kozhaya2024TRICK} that we aim to isolate and use for acquisition and, thus, for \ac{sop} navigation.

The algorithm used for beacon estimation assumes that the frame duration $T_{\mathrm{fr}}$ is constant, the data and pilot symbols are uncorrelated, and the power spectral density is stationary.
Let us consider a sampling period $T_{\rm s}$, and denote with $n = 0, 1, ... , N_{\mathrm{fr}} - 1$ the sample indices of the $k$-frame where $N_{\mathrm{fr}} = \mathrm{round}(T_{\mathrm{fr}}/T_\mathrm{s})$. Then, the digitized version of the received $k$-th frame after the baseband conversion can be expressed as

\begin{equation}\label{eq:signal_model}
    \bm{r}_k[n] = \bm{b}[n - d_k[n]]e^{j\theta_k[n]} + \bm{w}_k[n] \, ,
\end{equation}
where $\bm{b}$ is the beacon to be estimated, $d_k[n]$ is the time-varying code phase (in samples), $\theta_k[n]$ is the discretized phase variation across $T_\mathrm{fr}$, and $\bm{w}_k[n]$ is the \ac{awgn}. We remark that the noise term also collects the contribution due to information symbols transmitted within the generic $k$-th frame, which are typically zero-mean and can be averaged over the superposition of multiple frames.

Then, by approximating $d_k$ and $\theta_{k}$ via the Taylor series expansions of the zero and second order, respectively~\cite{Kozhaya2023Multi}, we can reformulate \eqref{eq:signal_model} as
\begin{equation}\label{eq:signal_model_approx}
\bm{r}_k[n] \approx \bm{b}[n - d_k]e^{j\bm{\Theta}_k[n]} + \bm{w}_k[n]\,,
\end{equation}
where 
\begin{equation}\label{eq:phase_Taylor}
    \bm{\Theta}_k[n] = \theta_k + \dot{\theta}_knT_{\rm s} + \frac{1}{2}\ddot{\theta}_k(nT_{\rm s})^2\,,
\end{equation} 
and $d_k$, $\theta_k$, $\dot{\theta}_k$, and $\ddot{\theta}_k$ denote the code and carrier phase and its first and second time derivatives, respectively, evaluated at the beginning of the $k$-th frame.

\section{Beacon Estimation Procedure Model}\label{sec:algorithm}

\begin{figure}
    \centering
    \includegraphics[width=1.0\linewidth]{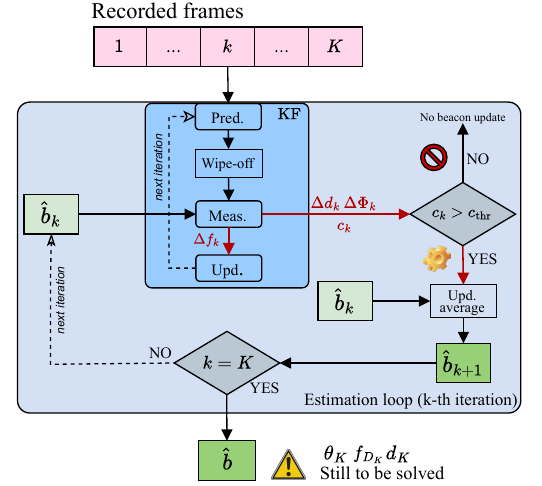}
    \caption{Working flow of the beacon estimator.}
    \label{fig:estimator_scheme}
    \vspace{-0.4cm}
\end{figure}



\begin{figure}
    \centering
    \includegraphics[width=1\linewidth]{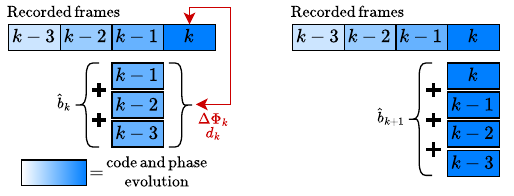}
    \caption{Forward beacon estimate update.}
    \label{fig:frame_forward}
    \vspace{-0.4cm}
\end{figure}

The beacon estimator is illustrated in Fig.~\ref{fig:estimator_scheme}. As can be seen, it operates iteratively over the sequence of $K$ recorded frames and is structured in a nested fashion. The inner loop consists of a \ac{kf} that tracks the carrier phase evolution across frames, enabling their coherent superposition over time. The outer loop establishes whether the just-obtained frame contains the pilot component and updates its estimate accordingly. 
In the next section, we detail the \ac{kf}-based estimator. 


\subsection{Kalman Filter for Phase Tracking}\label{subsec:kalman_filter}

The \ac{kf} used for estimation works in three steps, known as prediction, measurement, and update. Using the Taylor approximation \eqref{eq:phase_Taylor}, the \ac{kf} \emph{state} is $ \bm{x}_k \triangleq 
    \begin{bmatrix}
        \theta_k, &
        \dot{\theta}_k, &
        \ddot{\theta}_k
    \end{bmatrix}^\top$.
Note that $\dot{\theta}_k = 2\pi f_{{\rm D}_k}$, therefore the second and third components of the state vector correspond to the Doppler shift and Doppler shift rate, respectively, except for a factor $2\pi$.

Drawing again from \eqref{eq:phase_Taylor}, the state evolution is described by 
\begin{equation}
    \bm{x}_{k} = \bm{F} \bm{x}_{k-1} + \bm{\eta}\,,
\end{equation}
with $ \bm{\eta} \sim \mathcal{N}(0, \,\bm{Q})$, where 
\begin{equation}
    \bm{F} \triangleq 
\begin{bmatrix}
1 & T_{\mathrm{fr}} & \frac{1}{2}T_{\mathrm{fr}}^2\\
0 & 1 & T_{\mathrm{fr}} \\
0 & 0 & 1
\end{bmatrix}\, , \,
 \bm{Q} \triangleq q_w 
 \renewcommand*{\arraystretch}{1.5}
    \begin{bmatrix}
    \renewcommand*{\arraystretch}{1.5}
       \displaystyle {T_{\mathrm{fr}}^5}/{20} &  \displaystyle  {T_{\mathrm{fr}}^4}/{8}  &  \displaystyle {T_{\mathrm{fr}}^3}/{6} \\
 \displaystyle {T_{\mathrm{fr}}^4}/{8}  &  \displaystyle {T_{\mathrm{fr}}^3}/{3}  &  \displaystyle {T_{\mathrm{fr}}^2}/{2} \\
 \displaystyle {T_{\mathrm{fr}}^3}/{6}  &  \displaystyle {T_{\mathrm{fr}}^2}/{2}  &  \displaystyle T_{\mathrm{fr}}
    \end{bmatrix} \,.
\end{equation}
while, formally, the \emph{measurement model} is 
\begin{equation}\label{eq:measModel}
    z_k = \bm{H}\bm{x}_k + v\,,  \mbox{with}\, \bm{H} \triangleq
    \begin{bmatrix}
        0 & 1 & 0
    \end{bmatrix}
\end{equation}
where $ \bm{v} \sim \mathcal{N}(0, \, R)$.

After the \ac{kf} prediction step~\cite[Ch.\,13]{Kay1993Fundamentals}, a wipe-off phase compensation is applied to the frame according to \eqref{eq:phase_prediction}, as
\begin{equation}\label{eq:phase_prediction}
\bm{r}_{k}^c[n] = \bm{r}_{k}[n] \, 
e^{\, -j \left(
\theta_{k|k-1} + \dot{\theta}_{k|k-1}  n T_{\rm s} + \frac{1}{2} \ddot{\theta}_{k|k-1} (n T_{\rm s})^2 
\right)} 
\end{equation} 

The compensated frame is passed to the measurement block, whose working flow is detailed in Algorithm~\ref{alg:acquisition}. 
Note that the core of Algorithm~\ref{alg:acquisition} is a correlation via FFT between the compensated frame $\bm{r}_{k}^{\rm c}$, and the current beacon estimate $\hat{b}_k$. This provides their frequency offset $\Delta f_k$, as well as their offsets in terms of carrier phase $\Delta \Phi_k$ and code phase $\Delta d_k$. However, only $\Delta f_k$ is employed within the \ac{kf}, while $\Delta \Phi_k$ and $\Delta d_k$ are used outside the loop (Section~\ref{subsec:beacon_update}). 

During the update step, a traditional \ac{kf} compares prediction and measurement to compute the residuals~\cite[Ch.\,13]{Kay1993Fundamentals}. 
This \ac{kf} computes the residuals as $\tilde{y}_k = 2 \pi \Delta f_k$, where $\Delta f_k$ is obtained from the correlation step in Algorithm~\ref{alg:acquisition}. The a posteriori estimate is computed as in a traditional \ac{kf}, as 
\begin{equation}
  \bm{x}_{k|k}  = \bm{x}_{k|k-1} + \bm{K}_k\tilde{y}_k\,,
\end{equation}
where $ \bm{x}_{k|k-1} $ is a priori state estimate, obtained from the prediction and $\bm{K}_k$ is the Kalman gain~\cite[Ch.\,13]{Kay1993Fundamentals}.

It is important to highlight that the only a priori knowledge required for the \ac{kf} to converge and cancel out the contribution of \ac{awgn} and data is the frame duration $T_{\mathrm{fr}}$. Indeed, at the first iteration, when the process still has to start, $\hat{b}_1$ represents the first recorded frame as noisy as it was received.

The considered measurement model only makes use of the Doppler frequency shift, as shown in \eqref{eq:measModel}. 
Extended models using also code phase and derivative have already been used in the literature to extract the beacon from Orbcomm signals~\cite{kozhaya2022ORBCOMM}, to track Starlink signals~\cite{kassas_journal}, or to improve an already available beacon estimate~\cite{mooseli2025}. In turn, we used the model presented in~\cite{Kozhaya2023Multi} for its ease of implementation. Additionally, the Starlink code phase often has a non-deterministic behavior~\cite{qin2025}, thus it may be preferable to avoid its use when estimating the beacon from scratch. 

\begin{algorithm}
\caption{Beacon–Frame Correlation for Doppler, Phase, and Code Phase Measurement.}
\label{alg:acquisition}
\begin{algorithmic}

\REQUIRE $\hat{b}_k$, $r^{\rm c}_k$, $T_{\rm s}$, $N_\mathrm{fr}$, $f_{\rm min}$, $f_{\rm max}$, $f_{\mathrm{step}}$
\ENSURE $\Delta f_k$, $\Delta \Phi_k$, $\Delta d_k$, $c_k$ 

\STATE $f \leftarrow f_{\rm min} \mathpunct{:} f_{\mathrm{step}} \mathpunct{:} f_{\rm max}$
\STATE $\hat{B}_k \leftarrow \mathrm{FFT (\hat{b}_k)} $
\STATE $n_{T_{\rm s}} \leftarrow 0 : T_{\rm s} : (N_{\mathrm{fr}}-1)T_{\rm s}$
\STATE $\bm{C} \leftarrow \bm{0}_{|f| \times N_\mathrm{fr}}$

\FOR{$i = 1$ to $|f|$} 
    \STATE $f_i \leftarrow f(i)$ 
    \STATE$r^{\rm c}_{k, f_i}
\leftarrow r^{\rm c}_k \cdot \exp(-j 2 \pi f_i n_{T_{\rm s}})$
    \STATE $R^{\rm c}_{k, f_i} \leftarrow \mathrm{FFT}(r^{\rm c}_{k, f_i})$
    \STATE $C_i \leftarrow \hat{B}_k \cdot \mathrm{conj}(R^{\rm c}_{k, f_i})$
    \STATE $c_i \leftarrow \mathrm{IFFT}(C_i)$
    \STATE $\bm{C}(i,:) \leftarrow c_i$ 
\ENDFOR

\STATE $c_{k} \leftarrow \max(|\bm{C}|)$
\STATE $(i_\mathrm{max}, n_\mathrm{max}) \leftarrow \text{argmax}(|\bm{C}|)$
\STATE $\Delta f_k \leftarrow f(i_\mathrm{max})$
\STATE $\Delta d_k \leftarrow N_\mathrm{fr} - n_\mathrm{max}$
\STATE $\Delta \Phi_k \leftarrow -\angle \bm{C}(i_\mathrm{max}, n_\mathrm{max})$

\STATE \textbf{Return:} $\Delta f_k, \Delta \Phi_k, \Delta d_k, c_{k}$ 

\end{algorithmic}
\vspace{-0.1cm}
\end{algorithm}

\subsection{Beacon Update}\label{subsec:beacon_update}
First, notice that \ac{leo} transmissions are not continuous but depend on traffic demand on the ground. Thus, it may occur that the $k$-th processed frame contains only noise and no useful signal. In this case, updating the beacon estimate would degrade its quality and compromise subsequent compensations. For this reason, the refinement step is performed only if $c_k$ exceeds a predefined threshold $c_{\mathrm{thr}}$, which is empirically determined.

The carrier phase compensation \eqref{eq:phase_prediction} may not be sufficient to allow the fully coherent superposition of the frames, as only the residual Doppler shift is measured from the received signal. In turn, carrier and even code phases are not included in \eqref{eq:measModel} and are handled separately.
Then, $\hat{b}_k$ is aligned to the carrier phase and code phase of the current frame $r^{\rm c}_k$, as
\begin{equation}\label{eq:beacon_forward}
    \hat{b}^{c}_k = \mathrm{circshift}(\hat{b}_k, \Delta d_k) e^{j \Delta \Phi_k} \,,
\end{equation}
and the new beacon estimate will be
\begin{equation}\label{eq:beacon_average}
    \hat{b}_{k+1} = \frac{k}{k+1}\hat{b}^{c}_k + \frac{1}{k+1}r^c_k \,.
\end{equation}
This beacon refinement procedure is illustrated in Fig.~\ref{fig:frame_forward}, which shows that the current beacon is refined by adjusting its carrier phase and code phase to match the following frame. 

Once all $K$ iterations are completed, the estimated beacon is still affected by ambiguities on $\theta_K$, Doppler shift $f_{{\rm D}_{K}}$, and $d_K$. This occurs because the iterative loop measures the differences in carrier phase and code phase between consecutive frames and compensates them forward. Consequently, the corrections are propagated along the sequence, leading to a coherent superposition of $K$ frames that retains the residual ambiguities $\theta_K$, $f_{{\rm D}_{K}}$, and $d_K$ associated to the last $K$-th frame. The next Section details how to solve these ambiguities.

A crucial aspect for the success of the beacon estimation with a low-gain setup is the variability of the downlink transmission, which depends on the traffic demand of users in the satellite cell. As a result, the received signal may show fluctuations in power or duty cycle, making it non-trivial to identify portions of the signal suitable for estimation. Still, to increase the carrier-to-noise density ratio and, possibly, improve the beacon estimate, one may take advantage of \ac{tle} and \ac{sgp4} to select the best time instants for beacon estimation, e.g., when a \ac{sv} is close to the zenith.

\subsection{Ambiguities resolution} \label{sec:params_notes}

Compensating for the ambiguity $f_{{\rm D}_K}$ is fundamental, as an inaccurate value leads to a frequency bias in all subsequent acquired Doppler shifts. 
A first solution is obtained by inspecting the predicted Doppler shift, obtained from an orbital propagator, e.g., \ac{sgp4}. Despite its ease of implementation, this approach requires prior knowledge of the receiver's position and time. Another solution exploits the observed code phase to solve the ambiguity~\cite{Humphreys2023}.
The considered approach involves the correlation between \ac{pss} and \ac{sss} with the received signal to obtain a coarse estimate of $f_{D_K}$. 
Indeed, both methods can hardly guarantee the accuracy required for a precise estimation of $f_{D_K}$. Thus, the final value was empirically tuned by testing different candidate values. For instance, in the Starlink case, several $f_{D_K}$ hypotheses were evaluated until the demodulated \ac{ofdm} symbols showed the characteristic constellation of a \ac{psk} modulation in the IQ plane.

Concerning the code phase, $d_K$, its impact on the overall estimation quality depends on the \ac{sop} modulation itself.  For a time-native modulation, an incorrect frame alignment does not prevent the extraction of the beacon symbols, as it would only result in a circularly shifted version of the estimated beacon. Conversely, for \ac{ofdm}, as in Starlink, correct demodulation of the pilot symbols strictly requires the identification of the symbol boundaries, together with knowledge of the number of subcarriers and the cyclic prefix length. Results reported in~\cite{Humphreys2023} and~\cite{komodromos2023}, i.e., \ac{pss}, \ac{sss}, and \ac{ofdm} parameters, were exploited to determine $d_K$.

Finally, the estimate of $\theta_K$ has no impact for this study, since a constant phase offset applied to all symbols does not affect the Doppler shift estimates.

\begin{figure}
    \setlength{\fwidth}{.65\columnwidth}
    \input{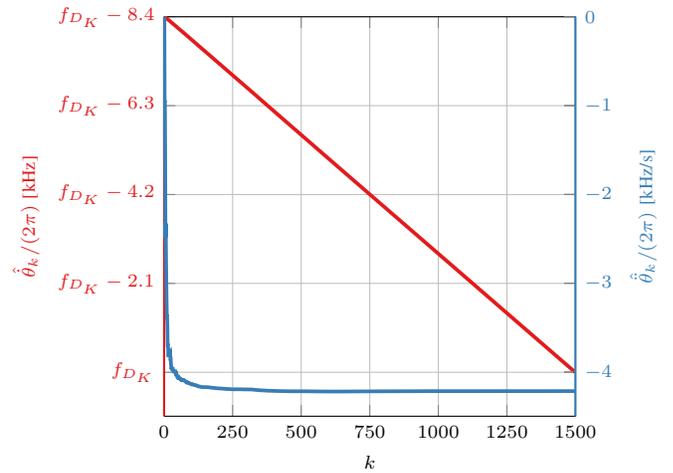}  
    \caption{Estimated Doppler and Doppler rate by the \ac{kf} as a function of the iteration.}
    \label{fig:dop_rate}
\end{figure}

\begin{figure}
    \input{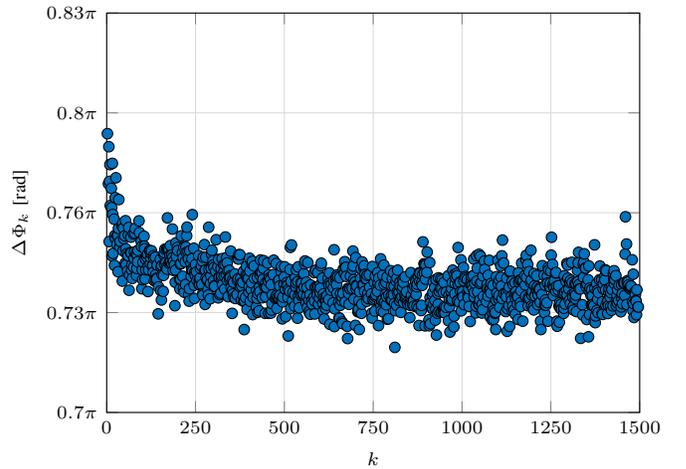}
    \caption{Phase differences between consecutive frames as a function of the iteration.}
    \label{fig:phase_differences}
    \vspace{-0.5cm}
\end{figure}

\begin{table}[t]
\centering
\renewcommand*{\arraystretch}{1.1}
\caption{Parameters for the Starlink beacon estimation.}
\label{tab:params}
\begin{tabular}{|>{\columncolor{paramCol}}c|>{\columncolor{valueCol}}c|
                >{\columncolor{paramCol}}c|>{\columncolor{valueCol}}c|}
\hline
\textbf{Parameter} & \textbf{Value} & \textbf{Parameter} & \textbf{Value} \\
\hline
$K$ & $1500~\mathrm{frames}$ 
& $\sigma_{\theta_0}$ & $0~\mathrm{rad}$ \\
\hline
$f_{\mathrm{min}}$ & $-10~\mathrm{Hz}$ 
& $\sigma_{\dot\theta_0}$ & $40\pi~\mathrm{rad/s}$ \\
\hline
$f_{\mathrm{max}}$ & $10~\mathrm{Hz}$ 
& $\sigma_{\ddot\theta_0}$ & $8000\pi~\mathrm{rad/s^2}$ \\
\hline
$f_{\mathrm{step}}$ & $0.5~\mathrm{Hz}$ 
& $R$ & $70~\mathrm{rad^2/s^2}$ \\
\hline
$c_{\mathrm{thr}}$ & $10^{9}$ 
& $q_w$ & $10^{-28}~\mathrm{rad^2/s^5}$ \\
\hline
$\theta_0$ & $0~\mathrm{rad}$ 
& $f_{D_K}$ & $29530~\mathrm{Hz}$ \\
\hline
$\dot{\theta}_0$ & $0~\mathrm{rad/s}$ 
& $\theta_K$ & $2.52~\mathrm{rad}$ \\
\hline
$\ddot{\theta}_0$ & $0~\mathrm{rad/s^2}$ 
& $d_K$ & $70966~\mathrm{samples}$ \\
\hline
\end{tabular}

\end{table}

\section{Starlink Beacon Estimation Results} \label{sec:stralink_beacon}

This section shows the results of the Starlink beacon estimation, considering both the estimation results and the analysis of the estimated beacon itself.

The estimated Doppler shift (as a function of the ambiguity $f_{{\rm D}_K}$) and Doppler rate over the $K$ iterations are shown in Fig.~\ref{fig:dop_rate}; note how the settling time of the Doppler rate is short compared to the overall duration of the process. Fig.~\ref{fig:phase_differences} illustrates the phase differences between the $k$-th beacon estimate and the corresponding frame, over all iterations satisfying $c_k\geq c_{\mathrm{thr}}$. 

Table~\ref{tab:params} reports the parameters set during the estimation. While many are capture-dependent, e.g., influenced by the specific hardware, these may be used to bootstrap the analysis.

\begin{figure}[t]
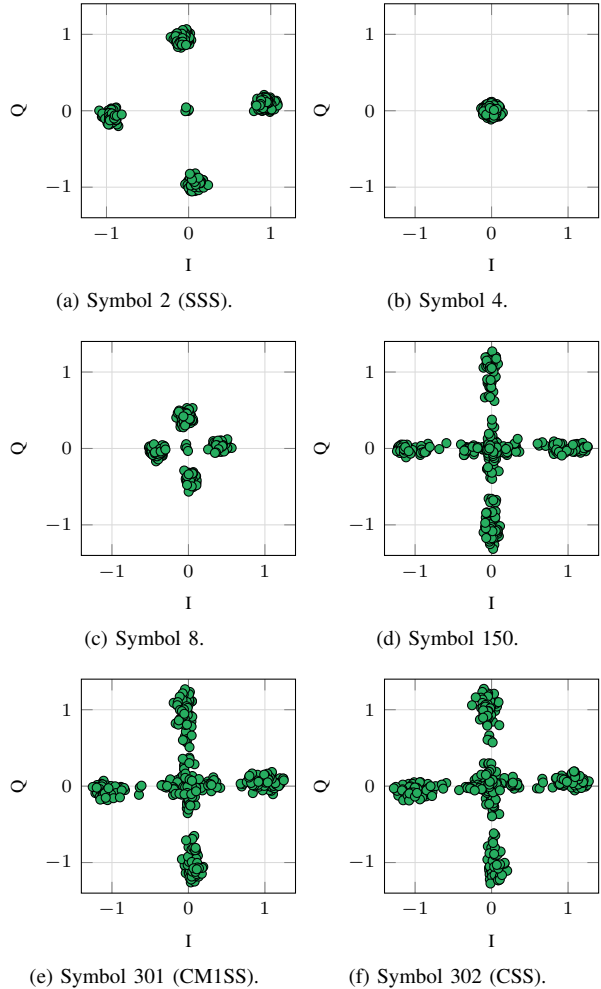

    \centering

    \setlength{\fheight}{0.5\columnwidth}
    \setlength{\fwidth}{0.5\columnwidth}

    \subfloat[Symbol 2 (SSS).\label{fig:symbol_2}]{\input{immagini/symbol2}}
    \subfloat[Symbol 4.]{\input{immagini/symbol4}}\\
    \subfloat[Symbol 8.]{\input{immagini/symbol8}}
    \subfloat[Symbol 150.\label{fig:symbol_150}]{\input{immagini/symbol150}}\\
    \subfloat[Symbol 301 (CM1SS).]{\input{immagini/symbol301}}
    \subfloat[Symbol 302 (CSS).]{\input{immagini/symbol302}}
    \caption{IQ plots for some of the \ac{ofdm} symbols in the Starlink frame after a normalization by the factor $10^4$.}
    \label{fig:symbols_grid}

    \vspace{-0.5cm}
\end{figure}

Once \(f_{{\rm D}_K}\), \(\theta_K\), and \(d_K\) are estimated, we can demodulate all the $302$ OFDM symbols composing the frame. Some of the demodulated symbols are shown in the IQ plane in Fig.~\ref{fig:symbols_grid}. A significant portion of the points lies close to \((1,0)\), \((0,1)\), \((-1,0)\), and \((0,-1)\), while the remaining are concentrated around the origin.

When coherently summing the \(K\) frames, being random, the transmitted information symbols should cancel out. Thus, only symbols with magnitude greater than \(0.5\) are considered recurrent and thus attributed to the Starlink beacon. 

Excluding \ac{pss} and \ac{sss}, inspection of the OFDM time--frequency grid shows that the detected pilot symbols constitute approximately $61.8 \%$ of the total transmitted symbols \footnote{Accounting only for the \SI{100}{\mega\hertz} front end.}. Moreover, almost the entire frame exhibits a regular pilot pattern, where one subcarrier every four is filled with pilot symbols for almost the  $T_{\mathrm{fr}}$ interval.
All OFDM symbols employ a 4-\ac{psk} modulation, and no phase jumps are observed across consecutive symbols, in contrast to what is reported in~\cite{Humphreys2023}.
 
Some of the estimated symbols exhibit slightly different characteristics with respect to the general behavior. For instance, the IQ plot of the \ac{sss} in Fig.~\ref{fig:symbol_2}, as it contains only pilot symbols and no information symbols across all subcarriers, shows only four points close to the origin, corresponding to the four gutter tones at the center of the channel, coherently with the frequency-domain description reported in~\cite{Humphreys2023}. In turn,  symbol 4 appears to be silent, while the 8-th seems to be entirely recurrent, similarly to the \ac{sss}, but transmitted at a lower power level.

\begin{figure}
    \input{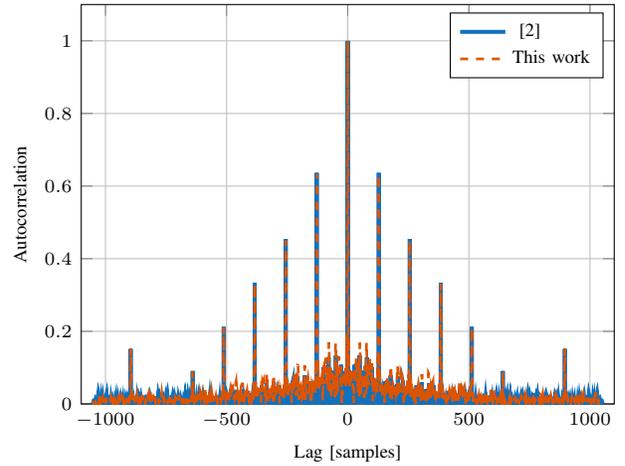}
    \captionsetup{aboveskip=0.01cm}
    \caption{Autocorrelation of the PSS: comparison between literature~\cite{Humphreys2023} and experimental results.}
    \label{fig:pss_autocorr}
    \vspace{-0.5cm}
\end{figure}

The \ac{pss} marks the beginning of the Starlink frame. This does not employ OFDM and is the repetition of the same subsequence, as shown in~\cite{Humphreys2023}. This peculiarity is clearly confirmed in Fig.~\ref{fig:pss_autocorr}, where the autocorrelation exhibits distinct peaks at non-zero lags.

\begin{figure}
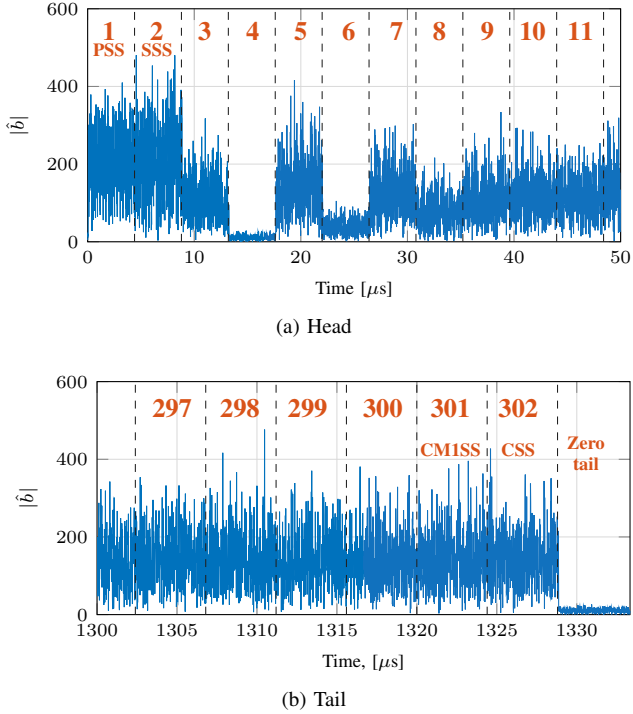

    \setlength{\fheight}{0.35\columnwidth}
     \centering
     \subfloat[Head]{\input{immagini/abs_starlink_beacon_head}}\\
     \subfloat[Tail]{\input{immagini/abs_starlink_beacon_tail}}
    \caption{Magnitude of the Starlink beacon: head (a) and tail (b).}
    \label{fig:frame_head_tail}
    \vspace{-0.5cm}
\end{figure}

Finally, the magnitude profile of the estimated Starlink beacon is shown in Fig.~\ref{fig:frame_head_tail}, which matches to the one in~\cite{Kozhaya2023Multi}.

The Starlink beacon obtained by applying the method described here to a signal capture taken in February 2025 is available at IEEE DataPort \footnote{https://dx.doi.org/10.21227/gmp1-gr28}. At the time of writing this paper (January 2026), the same estimate still seems to be valid for successfully acquiring measurements from new captures. 

\section{Acquisition and Navigation Framework} \label{sec:navigation}

\begin{figure*} 
    \centering
    \setlength{\fheight}{.6\columnwidth}
    \setlength{\fwidth}{\textwidth}
    \input{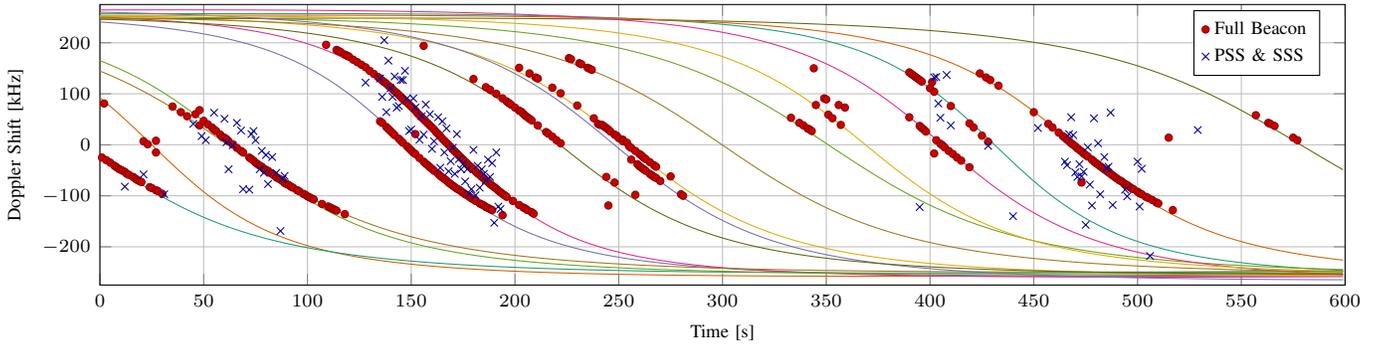}
        \caption{Starlink Doppler acquisition via PSS+SSS (blue crosses) and estimated beacon (red dots), using one frame. Solid lines indicate the expected Doppler shift from the orbital propagator for each satellite in view.}
\label{fig:confronto_doppler}
\end{figure*}

\subsection{Doppler Least Square Navigation Engine}\label{sec:DopLS}

We resort to a Doppler-based \ac{ls} to obtain the \ac{pvt} solution $\bm{s} = [\bm{p}, \bm{v}, \dot{\delta t}]^\top$, where $\dot{\delta t}$ is the user clock drift. We remark that the satellite clock bias does not impact the Doppler frequency shift, thus, it cannot be estimated from the considered measurements.
Since at each epoch the availability of enough measurements is not guaranteed, we perform \ac{pnt} using a set of measurements $\bm{f}_\mathrm{D} = [f_{{\rm D},0},\ldots,f_{{\rm D},L-1}]$, collected during a time window $\mathcal{T}$. 

Each measurement must be associated with an \ac{sv}, and thus with the \ac{sv} position and velocity.
However, unlike GNSS, where the satellite ID is identified during acquisition, when using LEO \ac{sop}, this information is not provided by the system. 
To do so, one may exploit the orbital propagator, set an initial position, and obtain the simulated Doppler curves (see Fig. \ref{fig:confronto_doppler}). Then, by comparing the estimations to their simulated counterpart, we can match each Doppler shift measurement to obtain the satellite ID. 
In this preliminary work, this matching is performed by associating each Doppler measurement with the closest predicted Doppler curve. Future work will focus on investigating alternative methodologies for this task and evaluating their benefits.
After this preliminary step, each measurement ${f_{D,\ell}}$, is now labeled with an \ac{sv} ID, and thus matched to a transmitter with position $\bm{p}_{\ell}$ and velocity $\bm{v}_{\ell}$. 

Now, we perform the actual iterative \ac{ls} to obtain the \ac{pvt} via iterative \ac{ls} from the Doppler frequency shifts, which, for the sake of convenience, are converted to range rate measurements, $\dot{r}_{\ell}$, as 
\begin{equation}
{f_{D,\ell}} =-\frac{f_{\rm c}}{c}  \frac{d{r}_{\ell}}{dt} = -\frac{f_{\rm c}}{c}  \dot {r}_{\ell}\,.
\end{equation}
First, we choose an initial solution $\bm{s}_0$.  
Then, at step $j$, for each solution $\bm{s}_j$, we can compute 
\begin{equation} \label{eq:pseudorate}
    \dot{r}_{{\ell},j} = \left(\bm{v}_{\ell} - \bm{v}_j\right)^\top \bm{e}_j  + c \dot{\delta t}\,, \,\mbox{with } \bm{e}_j=\frac{\bm{p}_{\ell} - \bm{p}_j}{\|\bm{p}_{\ell} - \bm{p}_j\|}\,.
\end{equation}

Next, we compute the residuals to build the system 
\begin{equation}\label{eq:LSmeasAndPos}
\Delta\dot{\bm{r}} = \bm{G} \Delta \bm{s}\,,\quad   \mbox{with}\quad    \Delta \dot{\bm{r}} = \dot{\bm{r}} - \dot{\bm{r}}_{j} - c\delta_j \bm{1}_L  \,,
\end{equation}
where $ \bm{1}_L$, is the all one column vector of size $L$, and $\bm{G} = \begin{bmatrix}
    \bm{G}_{\rm p} & \bm{G}_{\rm v}
\end{bmatrix} ^\top$
where each row in $\bm{G}$ is the Jacobian of \eqref{eq:pseudorate} computed at $\bm{s}_j$. In particular, the lines corresponding to each block are 
\begin{equation}
     \bm{G}_{{\rm p},{\ell}}=
         \frac{d\dot{\bm{r}}(\bm{s}_j)}{d\bm{p}} =   
              {\frac{(\bm{v}_{\ell}^\top\bm{e}_j) \bm{e}_j}{ \| \bm{p}_{\ell} - \bm{p}_j \| } - \frac{\bm{v}_{\ell} - \bm{v}_j}{ \| \bm{p}_{\ell} - \bm{p}_j \|  }}   \,,
\end{equation}
\begin{equation}
     \bm{G}_{{\rm v},{\ell}}=\begin{bmatrix}
         \frac{d\dot{\bm{r}}(\bm{s}_j)}{d\bm{v}},  &   \frac{d\dot{\bm{r}}(\bm{s}_j)}{d\dot{\delta t}} 
     \end{bmatrix} = \begin{bmatrix}
              \bm{e}_j,  & 1\end{bmatrix} \,.
\end{equation}
The solution update computed at each iteration is 
\begin{equation} \label{eq:pseudoinverse}
    \Delta \bm{s} = \left(\bm{G}^\mathrm{T}\bm{G}\right)^{-1} \bm{ G}^\top  \dot{\bm{r}}\,.
\end{equation}

Finally, the new solution is 
\begin{equation}
    \bm{s}_{j+1}  = \bm{s}_{j} + \Delta \bm{s}\,.
\end{equation}
The procedure is then iterated until convergence is reached, e.g., when $\|\Delta \bm{s}\| < \epsilon$, where $\epsilon$ is a predefined convergence threshold.
More details concerning the \ac{ls} can be found in~\cite[Ch.\,7]{kaplan2005understanding}, while examples of Doppler-aided \ac{ls} can be found in~\cite{Li2011Doppler,Shi2023Revisiting}.

We remark that, the \ac{ls}, could be replaced by more sophisticated approaches, such as a weighted \ac{ls}, \ac{kf}, or possibly even machine learning-based approaches. Still, in this preliminary work, we keep to this more straightforward approach, leaving the investigation on different solutions to future works.

Finally, a post-fit refinement is performed by discarding the measurement leading to large residuals, with respect to the current solution. The user set a priori a fraction $\lambda$ of the measurements to be kept. Then, by observing the empirical distribution of the post-fit residuals, the $1-\lambda$ fraction of the largest residuals is discarded. Finally, a new \ac{ls} is computed by using only the remaining measurements.
A trade-off shall be sought, as choosing a high $\lambda$ value allows us to reject most of the outliers but drastically reduces the number of available measurements.

\subsection{Positioning Results}\label{sec:positionResults}
Here, we report the final results obtained in terms of positioning performance, obtained using the Doppler shifts in Fig.~\ref{fig:confronto_doppler} as input. In this preliminary analysis, we set the orbital predictor to the true position $\bm{p}_{\mathrm{A}}$, i.e., the Qascom S.r.l headquarters in Cassola, Italy, to obtain the necessary data on the positions and velocities of the Starlink \acp{sv}. Then, two different \ac{ls} solution initializations have been tested: $\bm{p}_{\mathrm{A}}$ and $\bm{p}_\mathrm{B}$, i.e., the Department of Information Engineering (DEI) of the University of Padova, approximately $\SI{40}{\kilo\meter}$ away from $\bm{p}_{\mathrm{A}}$. 

When using the whole dataset, i.e., with $\mathcal{T}_1 = [0,600]\,$s and no refinement, the \ac{ls} converges to positions that are $\SI{1.06}{\kilo\meter}$ and $\SI{1.16}{\kilo\meter}$ away from the true receiver position when initializing the \ac{ls} to $\bm{p}_{\mathrm{A}}$ and $\bm{p}_{\mathrm{B}}$, respectively. This motivates the use of the post-fit refinement via thresholding, described in Section~\ref{sec:DopLS}.

\begin{figure}
    \centering
    \input{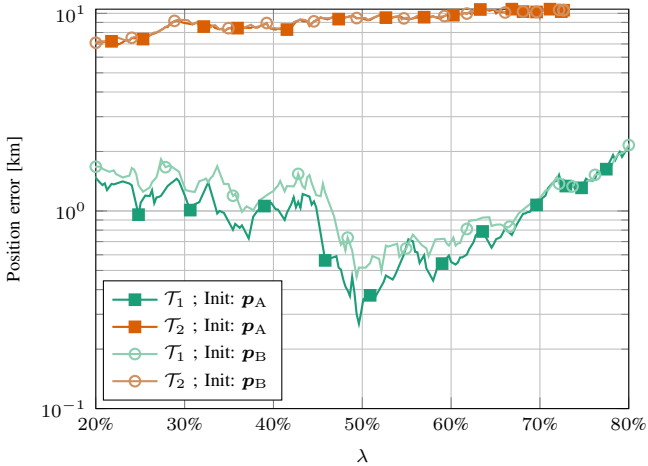}
    \caption{Position error vs threshold residual percentile for post-fit refinement, considering different initialization positions, and observation windows $\mathcal{T}$.}
    \label{fig:pos_err_residual}
    \vspace{-0.3cm}
\end{figure}

The curves in Fig.~\ref{fig:pos_err_residual} show the positioning error as a function of $\lambda$, considering the data collected in the time windows $\mathcal{T}_1$ and $\mathcal{T}_2 = [0,300]\,$s. The position error is minimized by setting $\lambda \approx 49\, \%$ for the measurement window $\mathcal{T}_1$, for both initializations. In particular, the position error is \SI{268}{\meter} when the \ac{ls} is initialized with the true position $\bm{p}_{\mathrm{A}}$, and \SI{464}{\meter} when the receiver was initialized in $\bm{p}_{\mathrm{B}}$. In turn, halving the observation window led to significantly worse results. In this case, the solution does not seem to be affected by \ac{ls} initialization.

These preliminary results provide several insights. First, since this rather simple refinement yields a significant improvement, more sophisticated techniques may yield even better results. Additionally, results show that it is better to start from a larger $\mathcal{T}$ and then perform a strict refinement than to start from a shorter $\mathcal{T}$ but keep a larger pool of measurements.

\section{Conclusions} \label{sec:conclusions}
The effectiveness with which \ac{sop} can be used for \ac{pnt} strongly depends on prior knowledge about the signal itself. In particular, taking advantage of recurring but undisclosed symbol sequences greatly improves the observables estimation accuracy. This paper demonstrates that the synchronization symbols employed by \ac{leo} constellations can be retrieved even with a lightweight setup, severely limited in front-end gain and bandwidth. In particular, we consider the Starlink signal as a case study. 
The implemented algorithms and the overall setup are validated by matching the estimated beacon to state-of-the-art results, such as the autocorrelation profile of the \ac{pss} and the beacon energy distribution.
Finally, we describe a Doppler frequency shift-based \ac{ls} that we use for \ac{pvt}, taking as input the measurement obtained via correlation with the \ac{sop} and the estimated beacon. To further improve the obtained results, we applied a post-fit refinement where we threshold measurement, leading to high measurement residuals. 
Testing is performed considering two observation windows and two solution initializations. An error of \SI{268}{\meter} is obtained by exploiting the entire capture duration and retaining $49 \,\%$ of the measurements corresponding to the lowest LS residuals.

\bibliographystyle{IEEEtran} 
\bibliography{IEEEabrv,references} 

\end{document}